# First observation of a mini-magnetosphere above a lunar magnetic anomaly using energetic neutral atoms

**Martin Wieser[1*], Stas Barabash[1], Yoshifumi Futaana[1], Mats Holmström[1], Anil Bhardwaj[2], R. Sridharan[2], M.B. Dhanya[2], Audrey Schaufelberger[3], Peter Wurz[3], Kazushi Asamura[4]**

[1] Swedish Institute of Space Physics, Box 812, SE-98128 Kiruna, Sweden

* email: wieser@irf.se

[2] Space Physics Laboratory, Vikram Sarabhai Space Center, Trivandrum 695 022, India

[3] Physikalisches Institut, University of Bern, Sidlerstrasse 5, CH-3012 Bern, Switzerland

[4] Institute of Space and Astronautical Science, 3-1-1 Yoshinodai, Sagamihara, Japan

**Abstract**

The Sub-keV Atom Reflecting Analyzer (SARA) instrument on the Indian Chandrayaan-1 spacecraft has produced for the first time an image of a lunar magnetic anomaly in backscattered hydrogen atoms. The image shows that a partial void of the solar wind, a mini-magnetosphere, is formed above the strong magnetic anomaly near the Crisium antipode. The mini-magnetosphere is 360 km across at the surface and is surrounded by a 300-km-thick region of enhanced plasma flux that results from the solar wind flowing around the mini-magnetosphere. The mini-magnetosphere is visible only in hydrogen atoms with energy exceeding 150 eV. Fluxes with energies below 100 eV do not show corresponding spatial variations. While the high-energy atoms result from the backscattering process, the origin of the low-energy component is puzzling. These observations reveal a new class of objects, mini-magnetospheres, and demonstrate a new observational technique to study airless bodies, imaging in backscattered neutral atoms.

## 1. Introduction

The lunar surface is directly exposed to solar wind plasma due to the Moon's lack of a magnetosphere or a dense atmosphere. This results in intense space weathering of the regolith covered surface (Hapke et al., 2001). When solar wind hits the surface, observations from lunar orbit show that a fraction of it is reflected as protons (Saito et al., 2008) and as neutral hydrogen atoms (Wieser et al., 2009). Although it lacks a global magnetic field, the Moon possesses regions of local magnetization, referred as magnetic anomalies, with magnetic field strengths of up to 100 nT at the surface (Mitchell et al., 2008). Using Lunar Prospector observations, Lin et al. (1998) suggested that the magnetic anomalies may create mini-magnetospheres, where the solar wind is deflected. Magnetohydrodynamic (MHD) simulations also predict the formation of mini-magnetospheres above strong magnetic anomalies (Harnett and Winglee, 2002). Energetic neutral atom imaging makes it possible to observe the presence of a mini-magnetosphere (Futaana et al., 2006): by shielding the surface from solar wind, a mini-magnetosphere produces a void in the observed flux of reflected neutral hydrogen atoms.

## 2. Instrumentation

The Sub-keV Atom Reflecting Analyzer (SARA) instrument (Bhardwaj et al., 2005, Barabash et al., 2009) on board the Indian Chandrayaan-1 spacecraft (Goswami and Annadurai, 2009), which orbited the Moon in a polar orbit, measured the energetic neutral atom flux from the lunar surface and simultaneously monitored the impinging flux of solar wind protons. The SARA instrument consists of two sensors, the Solar Wind Monitor (SWIM) and the Chandrayaan-1 Energetic Neutrals Analyzer (CENA). SWIM measures ions in the energy range from 10 eV to 15 keV with mass resolution (McCann et al., 2007); CENA measures energetic neutral atoms (10 eV to 3 keV) with moderate mass resolution within a 9° x 160° field of view (Kazama et al., 2007). Both sensors provide angular and energy resolution. CENA has a nadir-pointing field-of-view, whereas SWIM is directed partly toward the surface and partly toward space. Only hydrogen mass channels were used in this study. Figure 1 shows the orientation of the CENA field-of-view relative to the Moon and illustrates how the orbital motion of the spacecraft was used to produce the maps.

## 3. Observations

During nominal solar wind conditions, a large fraction of the impinging solar wind protons (up to 16% to 20%) is reflected back to space as energetic neutral hydrogen atoms (Wieser et al., 2009). Observations by SARA made on 17 June 2009 from an altitude of 200 km above the lunar surface show a reduction in the neutral atom flux from the surface above the strong magnetic anomaly at the Crisium antipode near the Gerasimovic crater (Hood et al., 2001) to less than half the value observed adjacent to the anomaly (Figure 2a). The reduction is most pronounced in the high-energy portion of the neutral atom energy spectrum, for energies between 150 eV and 600 eV. The region of reduced neutral atom flux has a diameter of about 360 km and is surrounded by a ring-shaped region about 300 km wide where enhanced energetic neutral hydrogen flux is observed. At lower energies (between 30 eV and 100 eV), the depletion at the center of the anomaly disappears, and the ring becomes a filled circle-like structure about 1000 km in diameter (Figure 2b). The reduced flux of neutral hydrogen is observed consistently during each orbit when passing over the magnetic anomaly, excluding temporal variation of the impinging solar wind. Average solar wind proton energy was 580 eV and alpha to proton ratio was less than 0.5% during these observations. Dynamic pressure of the solar wind plasma was between 1.0 nPa and 1.5 nPa.

The moon was outside the Earth's bow shock in undisturbed solar wind at a Geocentric Solar Ecliptic (GSE) longitude of 300° when the observations were made. Chandrayaan-1 was in an almost ideal noon-midnight orbit at this time. The images have been corrected for latitude-dependent solar wind input. No correction for a possible latitude-dependent angular emission of energetic neutral atoms from the surface was applied. We estimated that the total energetic neutral hydrogen flux in the undisturbed region is 20% of the solar wind flux. Throughout the observation interval, the interplanetary magnetic field was rather constant in the x-y plane in GSE coordinates with an azimuth of 125±10° and a magnitude of 5±1 nT based on magnetic field data from the Wind Magnetic Field Investigation (MFI) (Lepping et al., 1995).

Figure 3 shows energy spectra of the energetic neutral hydrogen flux at the center of the magnetic anomaly, of the surrounding ring with enhanced energetic neutral hydrogen flux, and of a region outside the ring. In contrast to the energetic neutral hydrogen flux, the proton flux from the Moon direction is strongly enhanced from the volume above the magnetic anomaly. The observed protons have mean energies of about 410 eV, which is slightly lower than solar wind proton energy of 580 eV. Increased proton flux is seen when the volume directly above the anomaly is within the field-of-view of surface-pointing SWIM pixels.

## 4. Discussion and Conclusions

Since the backscattered hydrogen flux is proportional to the impinging proton flux, the substantial reduction in the observed flux of energetic neutral atoms from the surface in correlation with the magnetic anomaly indicates effective shielding of the surface from solar wind. This is consistent with the predictions of Futaana et al. (2006). The close correlation between the reduction in neutral flux and the magnetic field data from Richmond and Hood (2008) indicates that a mini-magnetosphere is formed above the anomaly, deflecting solar wind.

The size of the mini-magnetosphere along the surface is about 360 km. The plasma flowing around the mini-magnetosphere results in increased ion flux onto the surrounding surface, resulting in an increased flux of reflected neutral atoms from within an annular region of about 300-km thickness. The indistinct outer boundary of this region of enhanced flux indicates that the formation of a bow shock is unlikely. The magnetic field within the anomaly is about 100 nT at the surface and about 20 nT at an altitude of 30 km (Mitchell et al., 2008; Richmond and Hood, 2008). The gyroradius of a 1-keV proton in this field is about 100 km. Therefore, the mini-magnetosphere is only two to three proton gyroradii across, and the enhanced flux region is three to four proton gyroradii across.

The mini-magnetosphere is formed similarly to a large-scale magnetosphere. The diamagnetic currents associated with the magnetic fields along the magnetopause deviate the solar wind plasma flow. It is surprising that this mechanism works even on such a small scale of a few gyroradii. Size and shape of the mini-magnetosphere are expected to strongly depend on solar wind conditions (Kurata et al, 2005). The solar wind dynamic pressure was rather low (< 1.5 nPa) during the observation interval, allowing the mini-magnetosphere to grow to the observed 360km diameter on the surface. Solar wind incident from near zenith direction is a likely cause for the tailless, spot-like shape of the reduced flux region. However, detailed numerical modeling is needed to establish the full three dimensional shape of the mini-magnetosphere.

The mini-magnetosphere is hardly visible in lower energy hydrogen atoms (< 100 eV), whereas it is pronounced in the energy range from 150 eV to 600 eV. This difference reveals the presence of two populations of hydrogen atoms that seem to have different origins. One population has lower energies, below 100 eV, while the other exhibits energies larger than 150 eV. The latter population is probably directly related to the impinging solar wind protons, because the mini-magnetosphere is clearly visible in images produced from these hydrogen atoms. The origin of the lower energy population, which does not seem to be affected by the mini-magnetosphere, is puzzling. To generate it, solar wind protons would need to be decelerated inside of the mini-magnetosphere, prior interaction with the surface. The observed neutral hydrogen flux from inside the mini-magnetosphere may also partly consist of recoils generated by impact of alpha particles or other heavy solar wind ions (e.g. multiply charged oxygen ions) onto the surface. Because of their larger gyroradii these ions would be less affected by the magnetic anomaly. Pick-up ions from the lunar exosphere or ions generated through a self-pick-up process (Saito et al., 2008) could also play a role. These ions could have energies higher than solar wind protons and could therefore penetrate deeper into the mini-magnetosphere.

Differences between lower and higher energy images may also reflect energy dependent angular scattering properties of regolith surfaces, with higher energy scatter products possibly being more specularly reflected than lower energy scatter products. In such a case, a part of the flux of higher energy neutral hydrogen inside the mini-magnetosphere would be missed due to the observation geometry.

Our observations provide direct proof that mini-magnetospheres do exist. Such objects may also be formed around asteroids, or be created artificially (Winglee et al., 2000).

Imaging in backscattered hydrogen atoms has proven to be a very effective method for investigating plasma structures close to rocky surfaces. Previous ideas for imaging of the surface relied on neutral atoms being sputtered from the surface by impacting ions (Grande et al., 1997; Futaana et al., 2006). The main disadvantages of the latter approach are very low fluxes of sputtered atoms (Wurz et al., 2007) and usually lower instrument sensitivity for heavy atoms (> 4 amu). The use of backscattered hydrogen overcomes these problems. Futaana et al. (2006) predicted fluxes of about $2{\cdot}10^5$ to $4{\cdot}10^5$ $cm^{-2}$ $s^{-1}$ $sr^{-1}$ of sputtered neutrals integrated over energy levels above 10 eV, while the measured neutral hydrogen fluxes ranged from about $2{\cdot}10^6$ to $8{\cdot}10^6$ $cm^{-2}$ $s^{-1}$ $sr^{-1}$, i.e., greater than the predicted values by a factor of about ten. Using backscattered hydrogen will be particularly effective for imaging regions of solar wind precipitation on Mercury, where strong hydrogen fluxes due to larger solar wind flux will permit shorter exposure times allowing to understand that highly dynamic magnetospheric system (Lukyanov et al., 2004).

## Legends

**Figure 1**:

Observational geometry for CENA. Its seven viewing directions (five shown) form a fan-shaped, nadir-pointing field-of-view whose greatest extent is in the cross-track direction. Coverage for mapping the energetic neutral atom flux from the surface is obtained by using the orbital motion of the spacecraft (S/C), as indicated by the velocity vector (v).

**Figure 2**:

Spatial variation in energetic neutral hydrogen flux from the surface over the magnetic anomaly near 22°S and 240°E on the lunar farside, observed from 200 km altitude on 17 June 2009. The maps show a unit-less reflection coefficient: neutral hydrogen number flux integrated over the specified energy range divided by total energy integrated solar wind number flux and cosine of lunar latitude.

a) In the energy range from 150 eV to 600 eV, a reduction in neutral hydrogen flux of about 50% is seen within the area of the mini-magnetosphere (dotted circle) compared to the surrounding ring-shaped region of enhanced flux (dashed line). Black contours in the center show the total magnetic field at 30 km altitude obtained from Lunar Prospector data (Richmond and Hood, 2008), with lines for 5 nT, 15 nT and 25 nT.

b) For lower energies, between 30 eV and 100 eV, the large-scale depletion in the neutral hydrogen flux above the magnetic anomaly is replaced by small-scale fluctuations, which are due in part to a low instrument count rate. The region of enhanced flux becomes almost a filled circle (dashed line).

c) Context image taken from the Clementine grayscale albedo map (Eliason et al., 1997;

Eliason et al., 1999; Isbell et al., 1999), available online at http://www.mapaplanet.com/explorer/moon.html. The black outline shows the area where energetic neutral hydrogen data is available, white dots represent the spacecraft ground track. Dashed black rectangles indicate locations where energy spectra in figure 3 were taken: inside mini-magnetosphere (M), enhanced flux region (E) and undisturbed region (U).

**Figure 3:**

Energy spectra of energetic neutral hydrogen atoms: from the surface within the mini-magnetosphere (open squares), from the enhanced flux region around the mini-magnetosphere (open circles), and from an undisturbed region outside the enhanced flux region (open triangles). Locations where these spectra were taken are indicated in figure 2c. The solar wind energy spectrum (solid squares; note the different y-axis on the right) is shown for comparison. Mean solar wind proton energy during these observations was 580 eV.

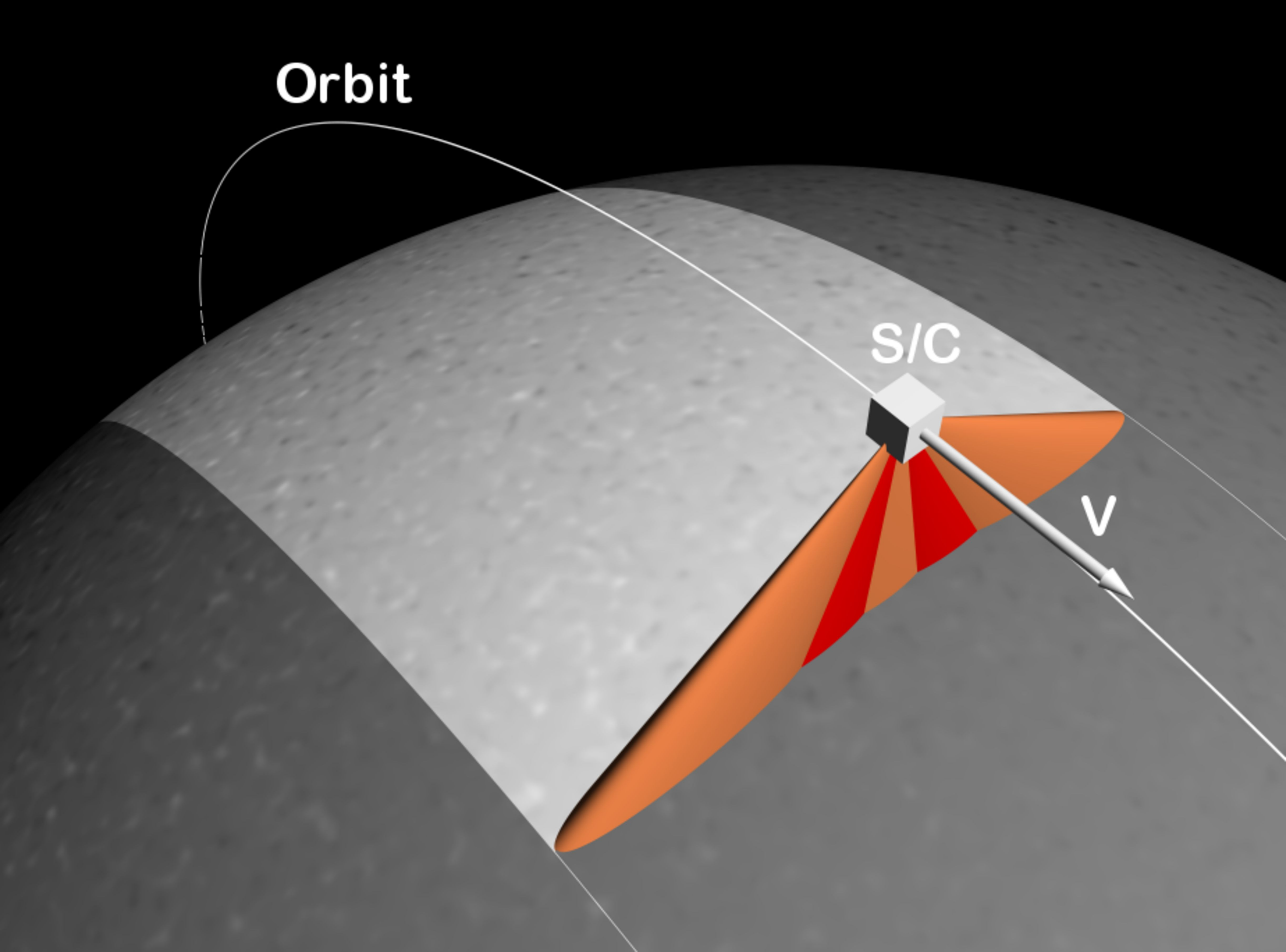
Orbit
S/C
V

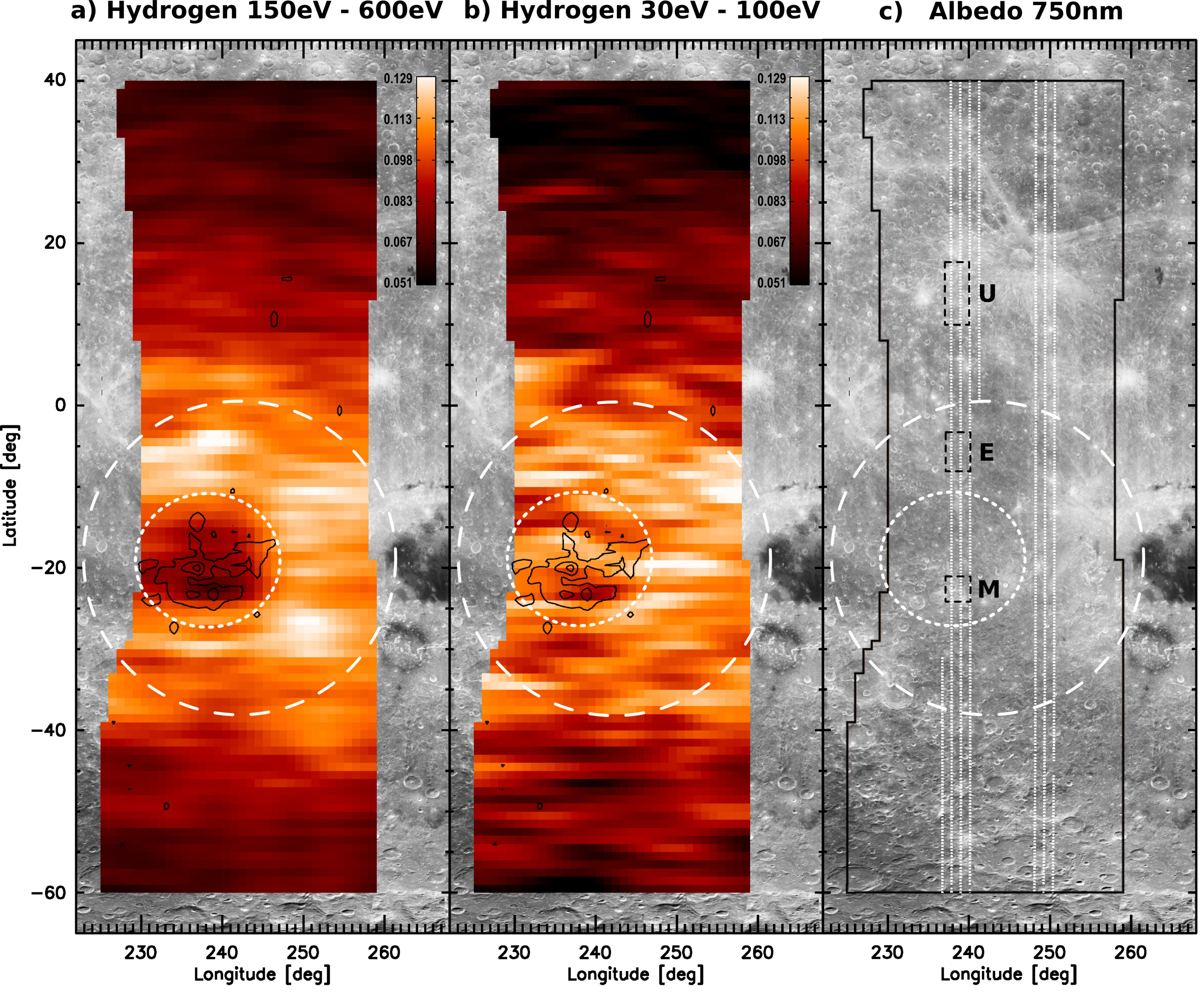

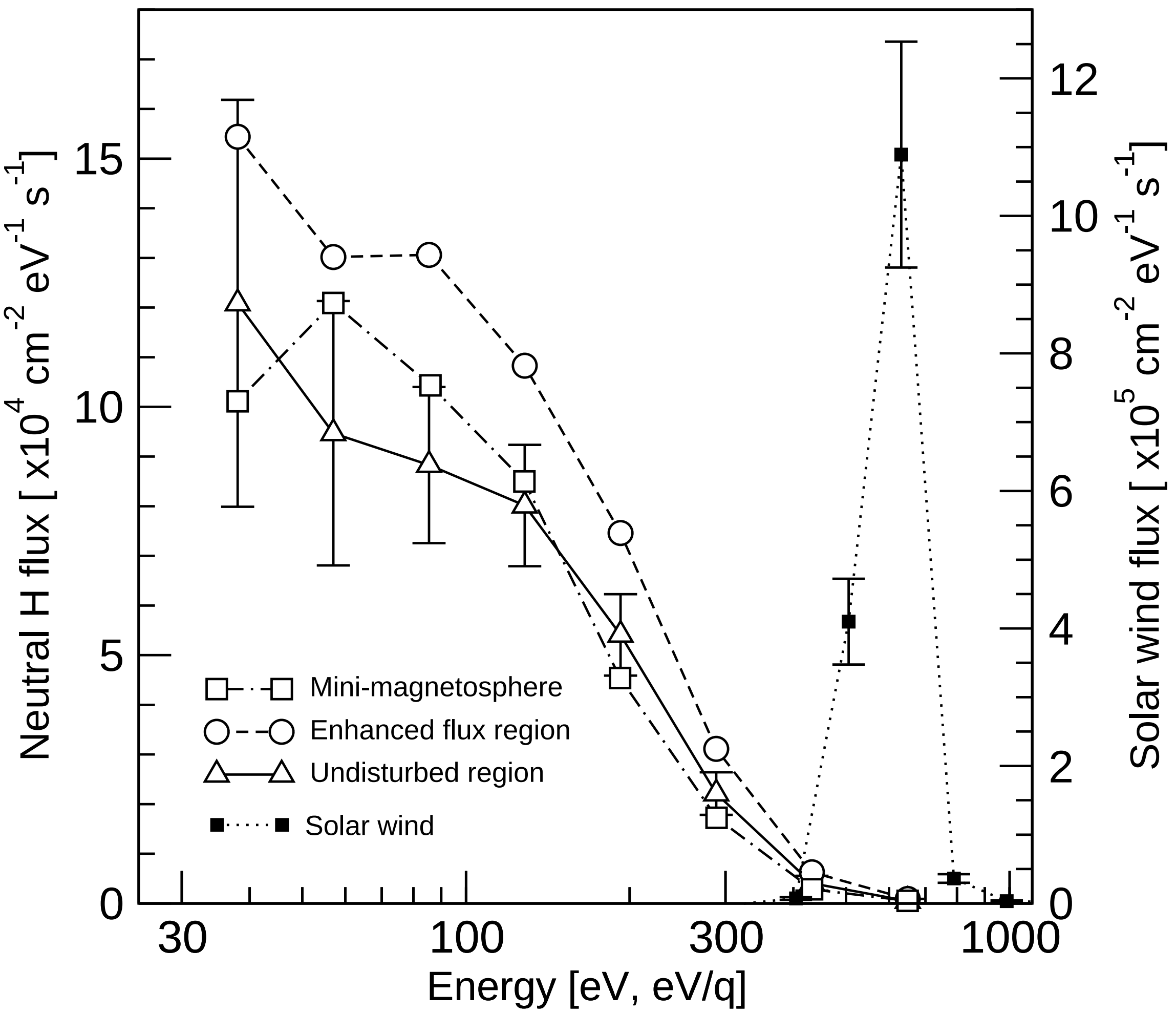

Neutral H flux [ x10$^4$ cm$^{-2}$ eV$^{-1}$ s$^{-1}$]
Solar wind flux [ x10$^5$ cm$^{-2}$ eV$^{-1}$ s$^{-1}$]
Energy [eV, eV/q]
Mini-magnetosphere
Enhanced flux region
Undisturbed region
Solar wind
0
5
10
15
2
4
6
8
10
12
30
100
300
1000